\documentclass[graybox]{svmult}

\usepackage{mathptmx}       
\usepackage{helvet}         
\usepackage{courier}        
\usepackage{type1cm}        
\usepackage{makeidx}         
\usepackage{graphicx}        
\usepackage{multicol}        
\usepackage[bottom]{footmisc}

\makeindex             

\usepackage[english]{babel}
\usepackage{amsmath}
\usepackage{amssymb}
\def\<{\langle}\def\>{\rangle}

\def\Supp{\mathsf{Supp}\,}
\def\tI{\transf{I}}

\def\tT{\transf{T}}

\def\rA{\sys{A}}\def\rB{\sys{B}}

\def\<{\langle}\def\>{\rangle}\def\kk{\>\! \>}\def\bb{\<\!\<}

\def\trnsfrm#1{\mathcal #1}

\def\tI{\trnsfrm I}\def\tT{\trnsfrm T}

\def\tP{\trnsfrm P } \def\tT{\trnsfrm T} 
\def\rA{{\rm A}}\def\rB{{\rm B}}

\def\St{\rm{St}}
\def\Trn{\mathrm{Transf}}

\def\Tr{\operatorname{Tr}}

\def\trnsfrm#1{\mathcal #1}

\def\tI{\trnsfrm I}\def\tT{\trnsfrm T}

\def\tP{\trnsfrm P } \def\tT{\trnsfrm T} 
\def\rA{{\rm A}}\def\rB{{\rm B}}

\def\sH{{\mathcal{H}}}\def\sS{{\mathcal{S}}}
\def\St{\rm{St}}\def\Trn{\mathrm{Trn}}\def\HS{\mathrm{HS}}
\def\Cmplx{\mathbb{C}}

\def\transp#1{{#1}^\intercal}
\def\eg{{e.~g.} }\def\ie{{i.~e.} }

\def\qed{$\,\blacksquare$\par}

\begin{document}

\title*{Quantum Holism\thanks{This work was made possible through the support of the Elvia and Federico Faggin Foundation, Grant 2020-214365.}}
\author{Giacomo Mauro D'Ariano}
\institute{Dipartimento di Fisica\\ dell'Universit\`a di Pavia, via Bassi 6, 27100 Pavia\\
              Tel.: +39 347 0329998   \email{dariano@unipv.it}\\
             \emph{Also:}  Istituto Lombardo Accademia di Scienze e Lettere\\ and
		INFN, Gruppo IV, Sezione di Pavia}
		%
%
\maketitle

\abstract{A composite quantum system has properties that are incompatible with every property of its parts. 
The existence of such global properties incompatible with all local properties constitutes what I call {\em mereological holism}--the distinctive holism of Quantum Theory. Mereological holism has the dramatic conceptual consequence of making untenable the usual understanding of  the "quantum system" as being a "physical object", since composed objects have properties compatible with those of its parts.\newline\indent
The notion of "property" can be extended in a unique way to the whole class of  operational probabilistic theories (shortly OPTs), of which the most relevant cases are Quantum Theory and Classical Theory. Whereas Classical Theory is not mereologically holistic, we can now search for other OPTs that are so. Within  the OPT framework the role of the "system" is that of an input-output connection between two objective events. In non holistic theories, such as Classical Theory, the system can still be regarded as an "object". On the contrary, in holistic theories interpreting "system" as "object" constitutes an hypostatization of a theoretical notion.}

\newpage
\section{Introduction}

What is a "physical object?"\footnote{Quine in his {\em Whither Physical Objects?} \cite{Quine} made a thorough attempt to arrive at a comprehensive meaning of  the word "object", but he ended up with a progressive evaporation of the concept--starting from that of  "body", toward "space-time region", ending with a mere "set of numerical coordinates".} 
In common language by it means a "thing", a tangible, visible, experienceable entity, which we identify with the bundle  of  its "properties".  An object without properties is inconceivable: it is not an object. Also, implicit in the notion of object,  is its mereological aspect, \ie its compositional nature in virtue of which putting objects together results in a new object, whose properties are made with the properties of its parts.  

\smallskip
In this paper it is shown that the common position of regarding  the quantum system as an "object" is untenable, since the quantum system has properties that are incompatible with any property of its parts (the "property" being described by an orthogonal projector over a subspaces of the system Hilbert space, as we learned from quantum logic \cite{BirkVN36}). Such part-whole incompatibility is ubiquitous in quantum theory,  as entanglement is.

The present result strongly supports the Copenhagen interpretation, according to which the quantum system (\ie  the "particle" in the non relativistic theory) is never directly observable, but only through its objective manifestations. On the contrary, in the common language the particle, as object, is  "something material that may be perceived by the senses" \cite{mw:object}.

The part-whole incompatibility reported here corresponds to a special notion of holism: the {\em mereological holism}. This is the distinctive holism of Quantum Theory. (The theory {\em is not} holistic in the only defined holism in the literature, namely the {\em local-discriminability} holism \cite{hw,CUPDCP}. 
A relevant conceptual consequence of mereological holism of Quantum Theory,  is that regarding the "system" as an "object" constitutes an hypostatization of a notion that is strictly theoretical, and whose role is just that of  {\em a theoretical input-output connection between objective events}, as we learnt from 
OPTs.\footnote{For an introductory review on OPTs, see Refs. \cite{CUPDCP,Chiribella2016}.} 
The "object" connotation of "system "can be recovered only in theories that are not mereologically holistic, such as Classical Theory. 
\medskip

After a short review about quantum properties in Section \ref{propertS}, the mereological holism of quantum theory is derived in Section \ref{noobj}.   In Section \ref{sectionqholism} it is remarked that the mereological holism is the distinctive quantum holism, and it is discussed its reconciliation with the of  local-discriminability.  In Section \ref{sectionop} the notion of "property" is univocally generalized to OPTs for further analysis of mereological holism in such theories. Section \ref{concl} concludes the paper summarizing the main results and proposing open problems and further analysis.

\section{Properties}\label{propertS}
Let's consider  a quantum system $\rA$ with associated Hilbert space $\sH_\rA$. To a proper subspace $\sS\subsetneq\sH_\rA$,  or, equivalently, to the corresponding orthogonal projector $P_\sS$,\footnote{There is trivially a one--to-one correspondence between Hilbert subspaces $\sS\subseteq\sH_\rA$ and orthogonal projectors $P_\sS$ on $\sS$.} we associate a {\em property} of the system $\rA$. Precisely, we say that
\begin{definition}\footnote{This notion of property has been introduced by John von Neumann in his celebrated treatise  \cite{von32}. Von Neumann noted that projections on a Hilbert space can be viewed as propositions about physical observables. The idea lead to the research line of {\em quantum logic}, the first major  attempt to derive Quantum Theory from first principles.}
The system $\rA$ in a state $\rho\in\St(\rA)$  with  $\Supp\rho\subseteq\sS$ has property $\sS$. If, instead,  $\Supp\rho\subseteq\sS^\perp$, then $\rA$ it has not the property $\sS$. When neither of the two cases apply the property $\sS$ is meaningless for state $\rho$.
\end{definition}
\begin{example}\label{EX1} Consider a two-dimensional system $\rA$ in the state corresponding to the vector $|\!\!\uparrow\>\in\Cmplx^2$. We say that the system has the property of  being "up", and it has not the property of  being "down"--the latter corresponding to state vector $|\!\!\downarrow\>$, with $\<\downarrow\!\!|\!\!\uparrow\>=0$. On the other hand, for state $|\!\!\rightarrow\>=
\tfrac{1}{\sqrt2} (|\!\!\uparrow\>+|\!\!\downarrow\>)$  properties "up" or "down" are  meaningless.
\end{example} 
\begin{example}\label{EX2} Consider a four dimensional system $\rA$ 
with canonical basis  $\{|n\>\}_{n=0}^4$. For  state $\rho=\tfrac{1}{4}|0\>\<0|+\tfrac{3}{4}|2\>\<2|$ we say that the system has the property of being {\em even}. The same property holds if  \eg   $\rho=\tfrac{1}{2}(|0\>+|2\>)(\<0|+\<2|)$.
\end{example} 

Thanks to the identifications {\em subspace $\equiv$ projector $\equiv$ property}, with a little abuse of language we will use the word {\em property} for all the three notions. 
\begin{example}[Property]
The state $\rho=|\!\!\!\uparrow\uparrow\>\<\uparrow\uparrow\!\!\!|\in\St(\rA\rB)$ has the property of being "symmetric", the property corresponding to the Hilbert subspace 
\begin{equation}\nonumber
(\sH_{\rA\rB})_+\subset\sH_{\rA\rB}\equiv\sH_{\rA}\otimes\sH_{\rB},
\end{equation}
$(\sH_{\rA\rB})_+$ being spanned by vectors invariant under the {\em swap} operator $E$, which acts on $\sH_{\rA\rB}$ as follows
\begin{equation}\label{swap}
E(|a\>\otimes|b\>)=|b\>\otimes|a\>.
\end{equation}
\end{example}
\begin{definition}
A property $P$ is nontrivial iff $0<P<I$.
\end{definition}
\par\medskip
Resorting to the usual notion of {\em compatibility} for observables, we say:
\begin{definition}
Two properties are compatible if and only if  they commute.
\end{definition}
\begin{example}[Incompatible properties] Properties {\em up} and {\em left} are incompatible (see also Example \ref{EX1}).
\end{example}
\begin{corollary}
\begin{enumerate}\item[]
\item Two properties are compatible iff their product is a property.
\item Two properties are mutually exclusive when $PQ=QP=0$.
\item $\dim(\sH_\rA)$ is the maximum number of mutually exclusive properties of $\rA$.
\end{enumerate}
\end{corollary}

\section{The quantum system is not an object!}\label{noobj}
In the following we will consider a bipartite system, and we will use the following nomenclature.
\begin{definition}[Property of the parts]\label{partwhole} We call {\em property of the parts} any property of the composite system $\rA\rB$ of the form $P_{\rA\rB}=P_\rA\otimes P_\rB$,  meaning that system $\rA$ has the property $P_\rA$, and system $\rB$ has the property $P_\rB$.
\end{definition}
\begin{definition}[Property of the whole]\label{partwhole} We call {\em property of the whole} any joint property $P_{\rA\rB}$ of the composite system $\rA\rB$ which is not a property of the parts.
\end{definition}
Generalization of the above definitions to multipartite systems are straightforward.
\bigskip
\par We are now in position to technically express and prove holism of Quantum Theory.
\begin{theorem}[Quantum Holism (properties)]\label{holprop}
There exist properties of the whole that are incompatible with any nontrivial property of the parts.
\end{theorem}
\proof  For the sake of proving existence, we can restrict to the simplest nontrivial case of a couple of systems $\rA,\rB$ with $\dim(\sH_\rA)=\dim(\sH_\rB)=d\leq\infty$ (extensions to more general cases will be provided after the present proof .)
\par A nontrivial property of the parts is represented by either a factorized projector of the form
\begin{equation}\label{2P}
P_{parts}=P\otimes Q,\qquad   (P,Q)\in\Pi^2_d 
\end{equation}
where $\Pi^2_d $ are the couples of orthogonal projectors in dimension $d$, of which at least one is non trivial. For the property of the whole we consider the following one
\footnote{Here we are using the {\em double-ket} notation \cite{bellobs} (for a thorough treatment see the book \cite{CUPDCP}.) Chosen the orthonormal factorized canonical basis $\{|i\>\otimes|j\>\}$ for $\sH\otimes\sH$, we have the one-to-one  correspondence between vectors in $\sH\otimes\sH$ and operators on $\sH$ 
\begin{equation}\nonumber
|\Psi\kk:=\sum_{ij}\Psi_{ij}|i\>\otimes|j\>\;\longleftrightarrow\quad\Psi=\sum_{ij}\Psi_{ij}|i\>\<j|\in \HS(\Cmplx^d),
\end{equation}
where $\HS(\Cmplx^d)$ denotes Hilbert-Schmidt operators in dimensions $d$. One then can veryify the following identity
\begin{equation}\nonumber
(A\otimes B)|C\kk=|AC\transp{B}\kk,
\end{equation}
$\transp{B}$ denoting the {\em transposed operator} of $B$, namely the operator having the transposed matrix w.r.t. the canonical basis. Notice that \eg $\transp{(|\phi\>\<\psi|)}=|\phi^*\>\<\psi^*|$ where $|\psi^*\>$ is the vector $|\psi\>$ with complex-conjugated coefficients w.r.t. the canonical basis.} 
 \begin{equation}\nonumber
P_{whole}=|\Gamma\kk\bb\Gamma |,\qquad \Gamma\in\HS(\Cmplx^d)\text{ invertible},\;\Tr(\Gamma^\dag\Gamma)=1.
\end{equation}
Let's consider the case of Eq. \eqref{2P}. We want to prove that there exist a $\Gamma$ such that
\begin{equation}\nonumber
\Big[P\otimes Q,|\Gamma\kk\bb\Gamma |\Big]\neq 0, \qquad\forall P,Q\in \Pi^2_d.
\end{equation}
Using the following identity for self-adjoint operators $A$ and $B$
\begin{equation}
[A,B]=2i\Im(AB),
\end{equation}
one obtains
\begin{equation}\nonumber
\Big[P\otimes Q,|\Gamma\kk\bb\Gamma |\Big]=2i\Im
[(P\otimes Q)|\Gamma\kk\bb\Gamma |].
\end{equation}
One has
\begin{equation}\nonumber
(P\otimes Q)|\Gamma\kk\bb\Gamma|=
|P\Gamma \transp{Q}\kk\bb\Gamma|
\end{equation}
whose imaginary part is zero if and only if
\begin{equation}\nonumber
|P\Gamma\transp{Q}\kk\bb\Gamma|=|\Gamma\kk\bb
P\Gamma\transp{Q}|,         
\end{equation}
namely 
\begin{equation}\nonumber
|P\Gamma\transp{Q}\kk=|\Gamma\kk,
\end{equation}
\ie 
\begin{equation}\label{psigammaphi}
P\Gamma\transp{Q}=\Gamma,
\end{equation}
which implies $P=Q=I$, since $\Gamma$  is invertible, contradicting the hypothesis 
that at least one of the two projectors is non trivial. \qed

\medskip
The above theorem can be easily generalized to two quantum systems with different dimensions, by considering $\Gamma$ only left or right-invertible, depending on which of the two systems has larger dimension. Upon considering one or both of the two systems $\rA$ and $\rB$ as composite, say $\rA=\rA_1\rA_2$,  the theorem can be easily generalized to three, or to any larger  number of subsystems, since incompatibility with all properties of $\rA$ implies incompatibility with any factorized property of $\rA$.

We have established the existence of properties of the whole that are incompatible with all properties of its parts, for every set of unbounded dimension and finite partitioning.\footnote{For infinite partitioning, one needs to move to a von Neumann algebra setting \cite{murphy1990c*-algebras}.}  In the following we will call such properties {\em holistic properties}. 

\medskip
The holistic properties that we have found are rank-one. One may proceed constructing higher-rank holistic properties \eg    by taking any operator $\Lambda$ orthogonal to $\Gamma$ (\ie  with $\bb\Gamma|\Lambda\kk=0$) using a Gram Schmidt procedure, thus building a full lattice of compatible holistic properties. And then by unitary transformation of the holistic properties obtain a set of  holistic properties that is presumably dense within the set of  compatible properties. One then may argue that, similarly to what happens for non separable states, the holistic properties are ubiquitous in quantum theory (\ie  they make a set dense within the set of properties), as it happens for state-entanglement. It would also be interesting to see if quantum discord may give rise to mereological holism (see \eg   Refs. \cite{PhysRevLett.108.120502,PhysRevA.81.052318}). 

\medskip We then reach the conclusion stated in terms of the following statement.
\medskip
\par\noindent {\bf No object interpretation:} the "quantum system" lacks the  "object" interpretation, since it misses its mereological connotation of having parts compatible with the whole.
\par

We end this section by commenting that a kind of Schroedinger-picture version of the above statement is the customary argument that by taking the marginal of a pure entangled state one gets a mixed state. This means that, quantifying  the "knowledge of the state" in terms of its von Neumann entropy, one has
\begin{proposition} The knowledge of the whole does not imply the knowledge of the parts.
\end{proposition}




\section{Mereological holism is the distinctive quantum holism.}\label{sectionqholism}
We have seen that quantum theory is "holistic", in the sense that there are properties of the whole that are incompatible with all possible properties of the parts, thus violating the mereological connotation of the notion of  "object" as being made of parts compatible with the whole. This has lead us to conclude that the  interpretation of the quantum "system" as "object" is untenable,  in contrast with the classical case, where all properties are compatible, hence the system can still be interpreted as being an object. 

In the arena of OPTs we have already encountered another kind of holism \cite{hw}, related to the principle of local discriminability \cite{CUPDCP}. Different theories can have different degrees of such holism, depending on being  local, bilocal, and generally $n$-local--the larger $n$, the more holistic the theory is.  We will rename such holism {\em local-discriminability holism} (shortly: {\em ld-holism}), to distinguish it from the kind of holism considered here, which we will refer to as  {\em mereological holism}. 

We should remind that as for Classical Theory, Quantum Theory  {\em is not} ld-holistic. Instead, contrarily to Classical Theory, Quantum Theory is mereologically holistic.  
Indeed, the fact that Quantum Theory is not ld-holistic guarantees its {\em compatibility with the scientific reductionistic approach} based on local observations in the presence of holism \cite{CUPDCP,maurofirst4}, and the kind of holism is  precisely the mereological one established here. 

\section{Notion of "property" and mereological holism in OPTs}\label{sectionop}
Whereas the ld-holism of OPTs has been analyzed in the literature \cite{hw,CUPDCP,maurofirst4,Francux-fermionic-discrim,CTnolocdiscr,PhysRevA.101.042118}, the same has still to be done for the mereological one. Many questions can be raised. The first that comes to mind is: Does mereological holism hold for all theories with entanglement, \eg symplectic theories with entanglement that have classical systems \cite{CTnolocdiscr,PhysRevA.101.042118}?

The above considerations suggest to confront different OPTs in regards of mereological holism, also for deepening our understanding of the notions of {\em system} in OPTs.  For such purpose we need to generalize the notion of {\em property} to the OPT level. 

At first sight this looks unfeasible, due to the quantum nature of the quantum-logic notion of "property", in one-to-one correspondence with Hilbert subspaces of the system Hilbert  space. However, we  will see now that the quantum notion of  property is in one-to-one correspondence with a genuine OPT notion: the {\em repeatable atomic transformation},\footnote{We remind that a trasformation $\tT$ is atomic if it is not the coarse-graining of other transformations, namely it cannot be written as the sum of two transformations e.g. as $\tT=\tT_1+\tT_2$.} namely an atomic transformation $\tP\in\Trn(\rA)$ satisfying the identity
\begin{equation}\label{RR}
\tP\tP=\tP.
\end{equation}
We recall that by the same definition of  "transformation" within the OPT framework \cite{CUPDCP}, Eq. \eqref{RR} holds independently on the test to which  $\tP$ belongs, and independently on the circuit in which the test is embedded,  namely independently on $\tP$ input state, and on what is performed at the output of $\tP$, thus holding also for non causal OPTs. It is easy to see  that there is a one-to-one correspondence between quantum repeatable atomic transformations $\tP$ and its relative projector $P$, since they are mapped to each other as follows
\begin{equation}
\tP=P\cdot P,\quad P=\Tr_1[(\tP\otimes\tI)(E)], \label{HURRA}
\end{equation} 
$E$ denoting the swap operator in Eq. \eqref{swap}. The second identity in Eq. \eqref{HURRA} is derived as follows
\begin{equation*}
\begin{aligned}
\Tr_1[\tP\otimes\tI)(E)]&=\Tr_1[(P\otimes I)E(P\otimes I)]=
\Tr_1[(E(P\otimes P)]\\&=\Tr_1[E (P\otimes I)]P=\Tr_1[(I\otimes P)E]P=
P\Tr_1[E]P=P.
\end{aligned}
\end{equation*} 
Thanks to bijection \eqref{HURRA} we can now identify a property with the corresponding repeatable atomic transformation $\tP$. We can then define the OPT notion of property in terms of a repeatable atomic transformation, and we can test mereological holism on various OPTs. Particularly interesting would be the case of  quantum-like theories (\eg   Fermionic) and of simplicial theories with entanglement \cite{CTnolocdiscr,PhysRevA.101.042118}. Moreover, it would be interesting to explore the interplay with OPT notion of complementarity \cite{MAPunpub} and the relation with incompatibility of properties, also in in connection with the principle of no-information without disturbance. 
\section{Conclusions}\label{concl}
In this paper  the mereological holism has been introduced, and shown that it is the distinctive holism of quantum theory. The mereological holism corresponds to the existence of global properties of systems that are incompatible with all local properties, where  "properties" are identified with orthogonal projectors. It has also been shown how to extend uniquely the notion of property to the whole class of  OPTs.  We are now in position to establish which other OPTs are mereologically holistic. Due to mereological holism, the usual interpretation of  {\em quantum system} as a {\em physical object} with "properties" is untenable. Therefore, regarding the "system" as an "object" constitutes a hypostatization of the strictly-theoretical concept of  "system", which in OPTs represents only a theoretical connection between objective events. 

The above conclusions suggest testing  mereological holism on different OPTs, based on a unique extension of  the notion of "property" to a repeatable atomic transformation, as suggested here. The concept of system recovers its "object" connotation only in theories which are not mereologically holistic, such as Classical Theory, whereas it is still an open problem weather this holds also for classical theories with entanglement, such as those in Refs. \cite{CTnolocdiscr,PhysRevA.101.042118}.
\medskip
\par\noindent\smallskip
{\bf Acknowledgements} I thank Paolo Perinotti for noticing the relevance of atomicity of the OPT version of  "property". 

\bibliographystyle{unsrt} 
\bibliography{quantum-holism-Springer-Econophys.bbl}
\end{document}